# Swinging of two-dimensional solitons in harmonic and Bessel optical lattices

Yaroslav V. Kartashov,[1,2] Victor A. Vysloukh,[3] and Lluis Torner[1]

[1]*ICFO-Institut de Ciencies Fotoniques, and Department of Signal Theory and Communications, Universitat Politecnica de Catalunya, 08034 Barcelona, Spain*

[2]*Department of Physics, M. V. Lomonosov Moscow State University, 119899, Moscow, Russia*

[3]*Departamento de Fisica y Matematicas, Universidad de las Americas - Puebla, Santa Catarina Martir, 72820, Puebla, Cholula, Mexico*

We consider parametric amplification of two-dimensional spatial soliton swinging in longitudinally modulated harmonic and Bessel lattices in Kerr-type saturable medium. We show that soliton center oscillations along different axes in two-dimensional lattices are coupled, which give rise to a number of interesting propagation scenarios including periodic damping and excitation of soliton oscillations along perpendicular axes, selective amplification of soliton swinging along one of transverse axes and enhancement of soliton spiraling.

Solitons in optically induced lattices were predicted and experimentally observed in photorefractive crystals in one and two transverse dimensions [1-4]. In photorefractive materials harmonic lattices are usually formed by the interference pattern of several plane waves whose intensity and intersection angles define the lattice depth and period. Such kind of lattices may be used for engineering of systems with tunable discreteness since they can operate in both regimes of weak and strong coupling between neighboring sites depending on the depth and period of refractive index modulation. Analogously to their discrete counterparts [5] lattice solitons can be used for a number of practical applications including all-optical switching and power-dependent soliton steering [6].

Properties of single solitons and soliton complexes supported by one- and two-dimensional optical lattices are now well established [7-13]. Lately we addressed properties of solitons supported by radially symmetric Bessel lattices [14]. Such lattices could be photoinduced by nondiffracting zero-order Bessel beams and offer a lot of new opportunities including control of soliton interactions in different lattice rings and possibility to set solitons into controllable rotary motion.

It was demonstrated recently that the presence of shallow *longitudinal* modulation of linear refractive index profoundly affects properties of solitons trapped in the guiding channel of one-dimensional optical lattice [15,16]. In particular, parametric amplification of transverse swinging and amplitude oscillations of spatial solitons is possible under appropriate conditions. The former effect can be potentially used for controllable soliton steering and fine-tuning of soliton inclination angle at the output face of the crystal. Two-dimensional generalization of this technique is far from being trivial because of the presence of the second transverse dimension, hence more complicated soliton trajectories. Moreover, despite the fact that in the absence of longitudinal modulation harmonic and Bessel lattices can support stable two-dimensional solitons even in a cubic medium, the open question is whether two-dimensional soliton is sufficiently robust to survive under remarkable longitudinal modulation of the linear refractive index which is necessary for effective parametric amplification of the soliton swinging?

In this paper we show that a considerable parametric amplification of soliton swinging can be achieved in two-dimensional case when small nonlinearity saturation is taken into account. We have found that transverse oscillations of the soliton center are coupled even in the absence of longitudinal modulation. This coupling is strong if frequencies of soliton beam oscillations along both transverse axes coincide, otherwise it is weak and parametric amplification of soliton center swinging along a selected axis is possible. We discuss some potential practical applications of parametric amplification of soliton swinging.

Our analysis is based on the nonlinear Schrödinger equation describing propagation of a laser beam in a medium with focusing Kerr-type saturable nonlinearity and spatial modulation of refractive index along longitudinal and transverse directions:

$$i\frac{\partial q}{\partial \xi} = -\frac{1}{2}\left(\frac{\partial^2 q}{\partial \eta^2} + \frac{\partial^2 q}{\partial \zeta^2}\right) - \frac{q|q|^2}{1+S|q|^2} - pQ(\xi)R(\eta,\zeta)q. \tag{1}$$

Here $q(\eta,\zeta,\xi)$ is the slowly varying dimensionless complex amplitude of the light field, transverse $\eta,\zeta$ and longitudinal $\xi$ coordinates are scaled in terms of beam radius and diffraction length, respectively, $S$ is the saturation parameter, guiding parameter $p$ is proportional to the refractive index modulation depth in transverse direction, functions $R(\eta,\zeta)$ and $Q(\xi)$ describe transverse and longitudinal refractive index profiles. Further we suppose that longitudinal variation of refractive index is described by the harmonic function $Q(\xi) = 1 - \mu\cos(\Omega_\xi \xi)$, where parameter $\mu < 1$ and $\Omega_\xi$ is the spatial frequency of longitudinal refractive index modulation. We consider two types of transverse profiles of refractive index: the harmonic one with $R_\mathrm{h}(\eta,\zeta) = \cos(\Omega_\eta \eta)\cos(\Omega_\zeta \zeta)$, where $\Omega_\eta, \Omega_\zeta$ are transverse spatial modulation frequencies, and the Bessel one $R_\mathrm{b}(\eta,\zeta) = J_0[(2b_\mathrm{lin})^{1/2}r]$, where $r = (\eta^2 + \zeta^2)^{1/2}$ is the radius, and parameter $b_\mathrm{lin}$ is the corresponding scaling factor (see Fig. 1(a) and 1(b)). The depth of refractive index modulation is assumed to be small compared with the unperturbed refractive index, and is of the order of nonlinear contribution due to the Kerr effect. Longitudinal modulation is supposed to be weak and smooth that enables one to neglect the reflected wave. In practice the refractive index modulation in transverse direction can be induced optically in photorefractive crystals with several interfering plane waves [1-4] or with nondiffracting Bessel beams [14]. Longitudinal modulation can be created in such media with spatially periodic background illumination along $\xi$-axis. Though in the case of real photorefractive crystal the model equation describing soliton propagation would be more complicated than Eq. (1) since the refractive index profile will also depend on the level of saturation, we expect that simplified model (1) adequately describes main qualitative features of soliton swinging. Notice that the total energy flow

$$U = \int_{-\infty}^{\infty}\int_{-\infty}^{\infty} |q|^2 \, d\eta \, d\zeta \tag{2}$$

remains constant upon propagation.

In the absence of longitudinal refractive index modulation Eq. (1) possess soliton solutions [7-14]. Here we recall basic properties of fundamental solitons. Soliton solutions can be found in the form $q(\eta,\zeta,\xi) = w(\eta,\zeta)\exp(ib\xi)$, where $w(\eta,\zeta)$ is the real function, and $b$ is the propagation constant. Substitution of this expression into Eq. (1) yields:

$$\frac{1}{2}\left(\frac{\partial^2 w}{\partial \eta^2} + \frac{\partial^2 w}{\partial \zeta^2}\right) + \frac{w^3}{1+Sw^2} + pR(\eta,\zeta)w - bw = 0. \quad (3)$$

Mathematically, families of soliton solutions of Eq. (3) are defined by parameters: $p$, $S$, $b$ and transverse configuration of the lattice. Thus, fundamental soliton supported by the Bessel lattice is radially symmetric and the position of its intensity maximum coincide with the center of the lattice, while in harmonic lattices soliton can be supported by either guiding site of the lattice. Here we found corresponding soliton profiles numerically using the standard relaxation method. Typically we used discretization scheme with $1024 \times 1024$ points per soliton profile, the transverse step was set to $d\eta = d\zeta = 0.02$. Zero boundary conditions were implemented. The progressive iterations in relaxation method were carried out until the relative difference between profiles on two successive iterations decreases below $10^{-20}$. The accuracy of calculations was checked by doubling the number of points per profile as well as by expanding of computation window (for broad solitons). The energy flow of solitons supported by Bessel and harmonic lattices versus propagation constant is shown in Fig. 1(c) for different values of the saturation parameter. For convenience of comparison we selected the scaling factor $b_{\text{lin}}$ for Bessel lattice in such way that the first zero of Bessel lattice coincides with that of the harmonic one (Fig. 1(a) and 1(b)). Energy flows $U_{\text{b,h}}$ of solitons supported by lattices of both types grow monotonically with increase of $b$ that indicates soliton stability [14] for chosen lattice parameters. Dispersion curves $U_{\text{b}}(b)$ and $U_{\text{h}}(b)$ are quite similar and differ notably only near a lower cut-off for soliton existence where soliton spreads over many lattice sites and exact periodicity of harmonic lattice and decaying behavior of the tail of Bessel lattice play a crucial role. As one can see from Fig. 1(c) the cutoff for solitons supported by Bessel lattices is a bit lower than that for solitons in harmonic lattices.

While exact solitons whose intensity maximum position coincides with maximum of the lattice will propagate in a stable way without any distortions, the small transverse displacement or tilt of the input soliton with respect to the lattice causes oscillations of the beam center in transverse plane upon propagation. Further, for illustration of main propagation scenarios of solitons in modulated lattices we solve Eq. (1) with an input condition

$$q(\eta,\zeta,\xi=0) = w(\eta-\eta_0,\zeta-\zeta_0)\exp(i\alpha_\eta\eta + i\alpha_\zeta\zeta), \qquad (4)$$

where $w(\eta,\zeta)$ is the exact soliton solution, $\eta_0,\zeta_0$ are initial shifts along $\eta$ and $\zeta$ axes, and $\alpha_\eta,\alpha_\zeta$ are input angles.

To understand multidimensional dynamics of tilted or shifted soliton beams in optical lattices one can use an effective particle approach [15,16], based on equations of motion for integral coordinates of the beam center:

$$\begin{aligned}\frac{d^2}{d\xi^2}\langle\eta\rangle &= p\frac{Q(\xi)}{U}\int_{-\infty}^{\infty}\int_{-\infty}^{\infty}|q|^2\frac{\partial R}{\partial \eta}d\eta\,d\zeta,\\ \frac{d^2}{d\xi^2}\langle\zeta\rangle &= p\frac{Q(\xi)}{U}\int_{-\infty}^{\infty}\int_{-\infty}^{\infty}|q|^2\frac{\partial R}{\partial \zeta}d\eta\,d\zeta.\end{aligned} \qquad (5)$$

Here integral coordinates $\langle\eta\rangle = U^{-1}\int_{-\infty}^{\infty}\int_{-\infty}^{\infty}\eta|q|^2\,d\eta\,d\zeta$ and $\langle\zeta\rangle = U^{-1}\int_{-\infty}^{\infty}\int_{-\infty}^{\infty}\zeta|q|^2\,d\eta\,d\zeta$, and Eqs. (5) are derived in the limit of cubic nonlinearity at $S\to 0$. The approach requires the substitution of a trial expression for the beam profile in the right sides of Eqs. (5). We use Gaussian beam $|q(\eta,\zeta,\xi)| = q_0\exp[-\chi_\eta^2(\eta-\langle\eta\rangle)^2]\exp[-\chi_\zeta^2(\zeta-\langle\zeta\rangle)^2]$, where $\chi_\eta,\chi_\zeta$ are form-factors and $q_0$ is the amplitude. In the simplest case of harmonic lattice one gets:

$$\begin{aligned}\frac{d^2}{d\xi^2}\langle\eta\rangle + [1-\mu\cos(\Omega_\xi\xi)]W_\mathrm{G}\Omega_\eta\cos(\Omega_\zeta\langle\zeta\rangle)\sin(\Omega_\eta\langle\eta\rangle) &= 0,\\ \frac{d^2}{d\xi^2}\langle\zeta\rangle + [1-\mu\cos(\Omega_\xi\xi)]W_\mathrm{G}\Omega_\zeta\cos(\Omega_\eta\langle\eta\rangle)\sin(\Omega_\zeta\langle\zeta\rangle) &= 0.\end{aligned} \qquad (6)$$

Here parameter

$$W_{\rm G} = p\exp[-(\Omega_\eta^2/\chi_\eta^2 + \Omega_\zeta^2/\chi_\zeta^2)/4] \qquad (7)$$

depends on the ratio between the characteristic lattice and beam scales, as well as on the depth of the lattice. In optically induced lattices this parameter can be fine-tuned by changing the lattice depth. This enables to control effectively dynamics of soliton motion inside the lattice for the same input profiles. Notice that other trial expressions for beam profile lead to the very similar equations for soliton center coordinates. As one can see from Eqs. (6), there is a straightforward analogy between equations of soliton movement in harmonic lattice and equations of motion for coupled parametrically driven pendulums. For the simplest case $\mu = 0$ and small initial soliton center displacements along $\eta$ and $\zeta$ axes, oscillations in these directions are independent, almost periodic, and occur at the certain frequencies given by $W_{\rm G}^{1/2}\Omega_\eta$ and $W_{\rm G}^{1/2}\Omega_\zeta$, respectively, that can be termed frequencies of free oscillations. Notice that for the Bessel lattice the frequency of free oscillations $\Omega_0$ is unique because of the radial symmetry of the lattice. For the case of relatively narrow solitons ($b_{\rm lin}^{1/2} \ll \chi_\eta, \chi_\zeta$) this frequency can be roughly estimated as $\Omega_0 \approx (pb_{\rm lin})^{1/2}\exp[-b_{\rm lin}/4\chi_\eta^2]$, assuming that $\chi_\eta = \chi_\zeta$.

However even at $\mu = 0$ large-amplitude oscillations of the soliton center along $\eta$ and $\zeta$ axes become coupled. This is an essentially new feature of two-dimensional soliton swinging in optical lattices in comparison with this phenomenon in one-dimensional lattices [15]. If large-amplitude oscillations along $\eta$-axis occur approximately at the same frequency as small-amplitude ones along $\zeta$-axis (i.e. when $\Omega_\eta = \Omega_\zeta$) the parametric resonance arises. Such parametric-type interaction opens an opportunity to transform effectively large amplitude $\eta$-oscillations into oscillations along $\zeta$-axis (Fig. 2(a)). This process repeats periodically in $\xi$ and looks like an amplitude beatings. Notice that under appropriate conditions predictions of the effective particle model (Eqs. (6)) are in a reasonable agreement with results of direct integration of Eq. (1) with input conditions (4). Thus at $\mu = 0$, $\Omega_\eta = \Omega_\zeta = 1$, $p = 6$, $\eta_0 = 1$, $\zeta_0 = 0.1$, $\alpha_\eta = \alpha_\zeta = 0$ and for soliton beam corresponding to $b = 7$ and $S = 0.1$, the difference between oscillations beating length $L_{\rm b} \approx 47.9$ obtained from Eq. (1) and $L_{\rm b}$ obtained on the basis of Eq. (6) for the Gaussian input beam with the same energy flow is about 13%.

More complicated situation occurs in lattices with longitudinal refractive index modulation. In such lattices the soliton center starts to swing with exponentially growing amplitude provided that the first parametric resonance condition $\Omega_\xi \approx 2\Omega_0$ is satisfied (here $\Omega_0$ is the frequency of free oscillations). The simplest case corresponds to the absence of coupling between $\eta$- and $\zeta$-oscillations. This can be achieved when soliton is initially shifted along the only one of transverse axes and swinging occurs in one plane (Fig. 2(b)). Notice that growth of the oscillation amplitude leads to diminishing of the instantaneous frequency, and the system escapes from the condition of parametric resonance. This results in periodic in $\xi$ decay and growth of soliton oscillations (or beatings). In this case the effective particle approach also offers quite realistic estimate for beating length and maximal value of transverse displacement of the soliton center. For example, at $\mu = 0.1$, $\Omega_\eta = \Omega_\zeta = 1$, $p = 6$, $\eta_0 = 0.1$, $\zeta_0 = 0$, $\alpha_\eta = \alpha_\zeta = 0$ and for soliton with $b = 7$, $S = 0.1$ the relative difference in calculation of beating length in the frames of two approaches is around 11%, and accuracy of calculation of the maximal value of transverse displacement of the soliton center is around 2.5%.

Another opportunity to avoid the coupling between $\eta$- and $\zeta$-oscillations is related with "selective" amplification of oscillations in the only one of transverse directions. This becomes possible when the frequencies of free oscillations for orthogonal axes are different (as in harmonic lattice with $\Omega_\eta \neq \Omega_\zeta$) and parametric resonance condition is fulfilled along only one axis. Fig. 3(a) illustrates the process of the selective parametric amplification.

The most complicated situation occurs when process of parametric amplification is accompanied by coupling between large-amplitude oscillations along $\eta$- and $\zeta$-axes. To realize such regime we shifted input soliton along $\zeta$-axis and simultaneously tilted it along $\eta$-axis. In the absence of longitudinal modulation soliton follows closed elliptical (or circular in particular case) trajectory, thus performing steady spiraling. Parametric amplification results in growth of the radius of the spiral trajectory on the initial stage of propagation. Finally spiral trajectory transforms into zigzag one, which means that oscillations along $\eta$- and $\zeta$-axes, which were initially phase-shifted by $\pi/2$, become phase-matched (Fig. 3(b)). After this stage of propagation the system escapes from the parametric resonance condition and the trajectory transforms into the elliptical one.

The parametric amplification of two-dimensional soliton swinging can be effectively used for the detection of the very small displacement and/or tilt of the input beam. For instance, in the longitudinally modulated Bessel lattice the initial displacement less than one percent of the beam width might be amplified parametrically up to the beam width, as illustrated in Fig. 4(a) (the white circle shows the first zero of the Bessel lattice). Moreover the parametric amplification can be used for fine-tuning of output tilt angle, while selective amplification can be used to enhance soliton oscillations in desired direction.

The key result of this work is summarized in Fig. 4(b), showing the resonance curve for transverse oscillations of two-dimensional soliton in the longitudinally modulated Bessel lattice, i.e. dependence of the ratio between maximal $\delta_{\max}$ and input $\delta_0$ values of transverse soliton displacement $\delta = (\langle \eta \rangle^2 + \langle \zeta \rangle^2)^{1/2}$ on relative frequency detuning $\nu = (2\Omega_0 - \Omega_\xi)/2\Omega_0$. This dependence has the form of a classical asymmetric parametric resonance curve for oscillator with "soft" sin-type nonlinearity. The maximum value of parametric amplification is reached at small negative value of frequency detuning. It should be also mentioned that the resonance curve is relatively narrow that allows highly selective amplification. We also want to stress that for a fixed frequency of longitudinal modulation $\Omega_\xi$ the parametric resonance conditions can be achieved by tuning the lattice depth $p$, since frequency of free oscillations $\Omega_0$ depends on the lattice depth, as it follows from Eqs. (6) and (7). Notice that oscillations of two-dimensional soliton are accompanied by radiation, but its rate is substantially reduced with growth of nonlinearity saturation and soliton energy flow.

In conclusion, we showed that in harmonic and Bessel lattices imprinted in Kerr-type saturable medium it is possible to achieve considerable parametric-type amplification of soliton swinging in the guiding lattice channel. This effect may find its applications for controllable soliton steering, for detection of submicron beam displacement and extremely small misalignments.

# Figure captions

Figure 1. (a) Harmonic and (b) Bessel optical lattices. Harmonic lattice is shown for $\Omega_\eta = \Omega_\zeta = 1$, while Bessel lattice corresponds to $b_{\text{lin}} = 1.172$. (c) Energy flow versus propagation constant for solitons supported by Bessel ($U_{\text{b}}$) and harmonic ($U_{\text{h}}$) lattices at different values of saturation parameter and $p = 5$. All quantities are plotted in arbitrary dimensionless units.

Figure 2 (color online). (a) Transformation of soliton center oscillations along $\eta$-axis into oscillations along $\zeta$-axis in the harmonic lattice without longitudinal refractive index modulation at $\Omega_\eta = \Omega_\zeta = 1$ and $p = 5.8$. Input conditions are $\eta_0 = 0.8$, $\zeta_0 = 0.08$, $\alpha_\eta = \alpha_\zeta = 0$. (b) Resonant parametric amplification of soliton swinging along $\zeta$-axis in longitudinally modulated Bessel lattice at $b_{\text{lin}} = 1.172$, $p = 5$, $\mu = 0.1$. Input conditions are $\eta_0 = 0$, $\zeta_0 = 0.05$, $\alpha_\eta = \alpha_\zeta = 0$. In (a) and (b) soliton beams correspond to $b = 7$ and $S = 0.1$. All quantities are plotted in arbitrary dimensionless units.

Figure 3 (color online). (a) Selective resonant parametric amplification of soliton swinging along $\eta$-axis in longitudinally modulated harmonic lattice at $\Omega_\eta = 1$, $\Omega_\zeta = 1.5$, $p = 5.8$, $\mu = 0.2$. Input conditions are $\eta_0 = \zeta_0 = 0.01$, $\alpha_\eta = \alpha_\zeta = 0$. (b) Complex soliton center trajectory in longitudinally modulated Bessel lattice at $b_{\text{lin}} = 1.172$, $p = 5$, $\mu = 0.1$. Input conditions are $\eta_0 = 0$, $\zeta_0 = 0.05$, $\alpha_\eta = 0.1$, $\alpha_\zeta = 0$. In (a) and (b) soliton beams correspond to $b = 7$ and $S = 0.1$. All quantities are plotted in arbitrary dimensionless units.

Figure 4. (a) Snapshot images showing maximal soliton displacements in positive and negative directions of $\zeta$-axis in longitudinally

modulated Bessel lattice. Input conditions are $\eta_0 = 0$, $\zeta_0 = 0.05$, $\alpha_\eta = \alpha_\zeta = 0$. (b) Maximal soliton displacement in longitudinally modulated Bessel lattice versus detuning. Input conditions are $\eta_0 = 0$, $\zeta_0 = 0.05$, $\alpha_\eta = 0.1$, $\alpha_\zeta = 0$. In (a) and (b) $b_{\mathrm{lin}} = 1.172$, $p = 5$, $\mu = 0.1$, and soliton beams correspond to $b = 7$ and $S = 0.1$. All quantities are plotted in arbitrary dimensionless units.

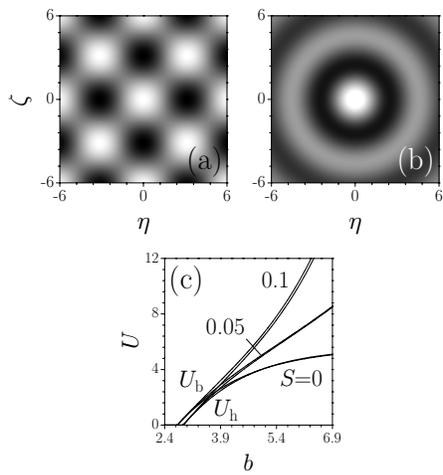

Figure 1.  (a) Harmonic and (b) Bessel optical lattices. Harmonic lattice is shown for $\Omega_\eta = \Omega_\zeta = 1$, while Bessel lattice corresponds to $b_{\mathrm{lin}} = 1.172$. (c) Energy flow versus propagation constant for solitons supported by Bessel ($U_{\mathrm{b}}$) and harmonic ($U_{\mathrm{h}}$) lattices at different values of saturation parameter and $p = 5$. All quantities are plotted in arbitrary dimensionless units.

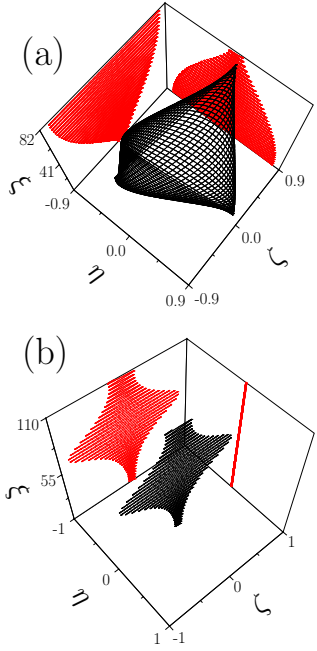

Figure 2 (color online). (a) Transformation of soliton center oscillations along $\eta$-axis into oscillations along $\zeta$-axis in the harmonic lattice without longitudinal refractive index modulation at $\Omega_\eta = \Omega_\zeta = 1$ and $p = 5.8$. Input conditions are $\eta_0 = 0.8$, $\zeta_0 = 0.08$, $\alpha_\eta = \alpha_\zeta = 0$. (b) Resonant parametric amplification of soliton swinging along $\zeta$-axis in longitudinally modulated Bessel lattice at $b_{\text{lin}} = 1.172$, $p = 5$, $\mu = 0.1$. Input conditions are $\eta_0 = 0$, $\zeta_0 = 0.05$, $\alpha_\eta = \alpha_\zeta = 0$. In (a) and (b) soliton beams correspond to $b = 7$ and $S = 0.1$. All quantities are plotted in arbitrary dimensionless units.

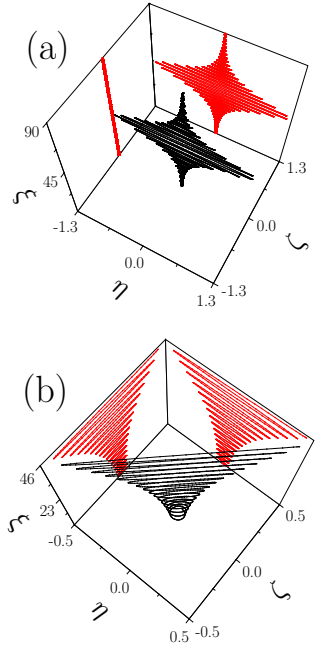

Figure 3 (color online). (a) Selective resonant parametric amplification of soliton swinging along $\eta$-axis in longitudinally modulated harmonic lattice at $\Omega_\eta = 1$, $\Omega_\zeta = 1.5$, $p = 5.8$, $\mu = 0.2$. Input conditions are $\eta_0 = \zeta_0 = 0.01$, $\alpha_\eta = \alpha_\zeta = 0$. (b) Complex soliton center trajectory in longitudinally modulated Bessel lattice at $b_{\mathrm{lin}} = 1.172$, $p = 5$, $\mu = 0.1$. Input conditions are $\eta_0 = 0$, $\zeta_0 = 0.05$, $\alpha_\eta = 0.1$, $\alpha_\zeta = 0$. In (a) and (b) soliton beams correspond to $b = 7$ and $S = 0.1$. All quantities are plotted in arbitrary dimensionless units.

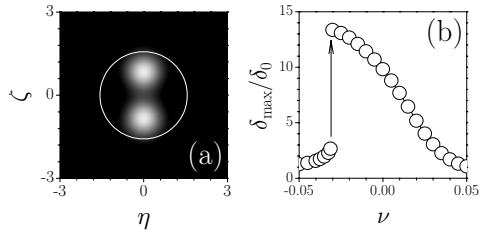

Figure 4. (a) Snapshot images showing maximal soliton displacements in positive and negative directions of $\zeta$-axis in longitudinally modulated Bessel lattice. Input conditions are $\eta_0 = 0$, $\zeta_0 = 0.05$, $\alpha_\eta = \alpha_\zeta = 0$. (b) Maximal soliton displacement in longitudinally modulated Bessel lattice versus detuning. Input conditions are $\eta_0 = 0$, $\zeta_0 = 0.05$, $\alpha_\eta = 0.1$, $\alpha_\zeta = 0$. In (a) and (b) $b_{\text{lin}} = 1.172$, $p = 5$, $\mu = 0.1$, and soliton beams correspond to $b = 7$ and $S = 0.1$. All quantities are plotted in arbitrary dimensionless units.